# Dynamic Warp Resizing in High-Performance SIMT

Ahmad Lashgar[1]
a.lashgar@ece.ut.ac.ir

Amirali Baniasadi[2]
amirali@ece.uvic.ca

Ahmad Khonsari[1 3]
ak@ipm.ir

[1]School of ECE
University of Tehran

[2]ECE Department
University of Victoria

[3]School of Computer Science
Institute for Research in
Fundamental Sciences

*Abstract*—Modern GPUs synchronize threads grouped in a warp at every instruction. These results in improving SIMD efficiency and makes sharing fetch and decode resources possible. The number of threads included in each warp (or warp size) affects divergence, synchronization overhead and the efficiency of memory access coalescing. Small warps reduce the performance penalty associated with branch and memory divergence at the expense of a reduction in memory coalescing. Large warps enhance memory coalescing significantly but also increase branch and memory divergence. Dynamic workload behavior, including branch/memory divergence and coalescing, is an important factor in determining the warp size returning best performance.

Optimal warp size can vary from one workload to another or from one program phase to the next. Based on this observation, we propose Dynamic Warp Resizing (DWR). DWR takes innovative microarchitectural steps to adjust warp size during runtime and according to program characteristics. DWR outperforms static warp size decisions, up to 1.7X to 2.28X, while imposing less than 1% area overhead. We investigate various alternative configurations and show that DWR performs better for narrower SIMD and larger caches.

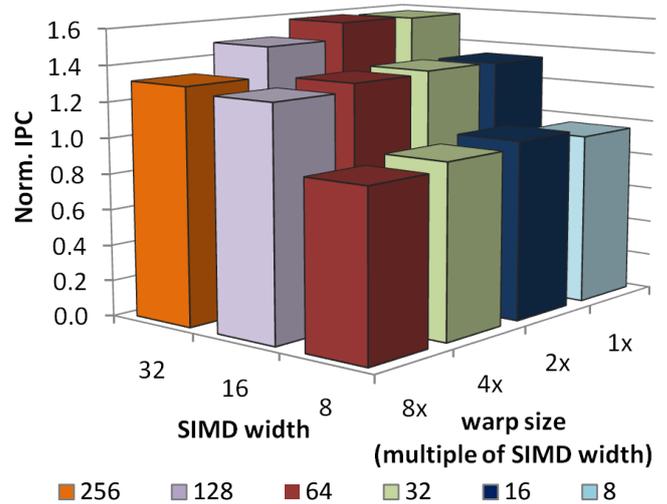

**Figure 1. Warp size impact on performance for different SIMD widths, normalized to 8-wide SIMD and 2x warp size.**

## I. INTRODUCTION

Conventional SIMT accelerators achieve high performance by executing thousands of threads concurrently. In order to maintain design simplicity, neighbor threads are bundled in groups referred to as warps. Employing warp-level granularity simplifies the thread scheduler as it uses coarse-grained schedulable elements. In addition, this approach keeps many threads at the same pace providing an opportunity to exploit common control-flow and memory access patterns. Memory accesses of neighbor threads within a warp can be coalesced to reduce the number of off-core requests. The underlying SIMD units are more efficiently utilized as a result of executing warps built using threads with the same program counter and behavior. Parallel warps amortize the communication overhead associated with waiting threads by using computations required by other threads.

GPUs are still far behind their potential peak performance as they face two important challenges: branch and memory divergence [9]. Upon branch divergence, threads at one side of a branch stay active while the other side becomes idle. Upon memory divergence, threads hitting in cache have to wait for those who miss. At both divergences, threads suffer from unnecessary waiting periods. This waiting can result in performance loss as it can leave the core idle.

One of the parameters strongly affecting the performance impact of such divergences is the number of threads in a warp or warp size. *Small warps*, i.e., warps as wide as SIMD width, reduce the likelihood of branch/memory divergence occurrence. Reducing branch divergence reduces the number of inactive-threads at diverging paths and waiting-threads at re-convergence point. Moreover, reducing memory divergence reduces unnecessary waiting imposed to hit threads. On the other hand, small warps reduce memory coalescing, which can increase memory stalls. This can lead to redundant memory accesses and increase pressure on the memory subsystem. *Large warps,* on the other hand, exploit potentially existing memory access localities among neighbor threads and coalesce them to a few off-core requests. On the negative side, large warps can increase serialization and the branch/memory divergence frequency.

Figure 1 reports average performance for benchmarks used in this study (see methodology for details) for different warp sizes and SIMD widths. For any specific SIMD width, configuring the warp size to 1-2X larger than SIMD width provides best average performance. Widening the warp size beyond 2X degrades performance. In the remainder of this paper, we use an 8-wide SIMD configuration.

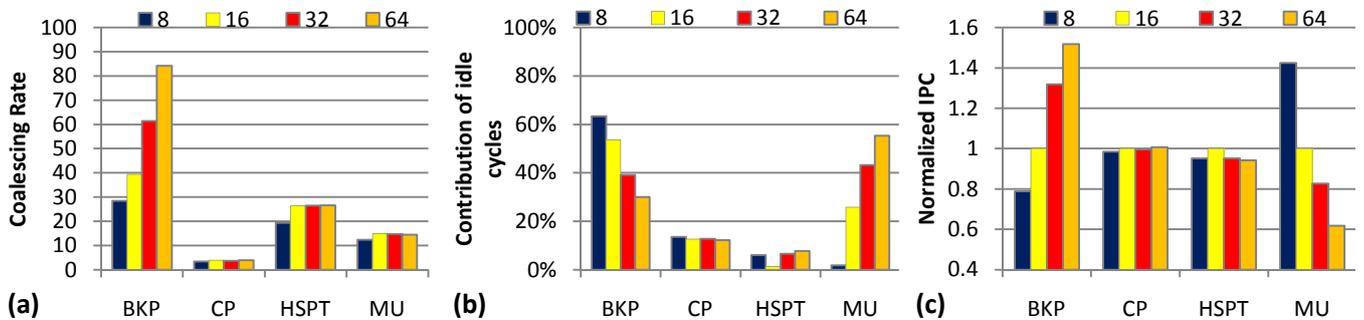

Figure 2. (a) Coalescing rate, (b) Idle cycle share and (c) Performance under different warp sizes. IPC is normalized to a GPU using 16 threads per warp.

In this paper we analyze how warp size impacts performance in GPUs. We start with studying GPUs using different warp sizes. We use our analysis and introduce Dynamic Warp Resizing (DWR) to achieve both coalescing benefits (associated with large warps) and low synchronization overhead (associated with small warps).

In summary we make following contributions:

- We evaluate the effect of warp size on GPU performance under general-purpose workloads. We also investigate warp size impact on coalescing rate, and idle cycles.

- We introduce DWR to achieve performance benefits of both small and large warps. We do so by adjusting warp size dynamically and according to program behavior.

- We propose a realistic hardware implementation for DWR and evaluate the associated overhead.

- We evaluate DWR under various microarchitectures, including those with different SIMD width, and L1 cache size.

The rest of the paper is organized as follows. In Section II we study background. In Section III we review warp size impact. In Section IV we present DWR. In Section V we discuss methodology. Section VI reports results. In Section VII we discuss our findings in more detail. In Section VIII we review related work. Finally, Section IX offers concluding remarks.

## II. BACKGROUND

In this study we focus on SIMT accelerators similar to NVIDIA Tesla [1] [10]. Stream Multiprocessors (SMs) are processing cores and send memory requests to memory controllers through on-chip crossbar network. We augment Tesla with private L1 caches for each SM.

Each SM keeps context for 1024 threads. While recent GPUs (e.g. NVIDIA Kepler [16]) have multiple warp schedulers issuing instructions on multiple SIMD groups, Tesla's SM has one thread scheduler which groups and issues warps on one SIMD group. Threads within a warp have the same program counter. Control-flow divergence among threads is managed using re-convergence stack [3, 5] where diverged

[1] In this study Tesla refers to the Tesla architecture not Tesla graphic card brand.

threads are executed serially until re-converging at the immediate post-dominator.

Instructions from different warps are issued back-to-back in a 24-stage, 8-wide SIMD pipeline. In the absence of ready warps in the warp pool, the pipeline front-end stays idle leading to underutilization. A significant portion of such underutilization periods could be eliminated by executing threads, which are ready yet inactive/waiting due to branch/memory divergence [11].

In this work we model a coalescing behavior similar to compute compatibility 2.0 [15]. Requests from neighbor threads accessing the same stride are coalesced into one request. Consequently, memory accesses of a warp are coalesced into one or many stride accesses. Each stride is 64 bytes. Our memory transaction granularity is equal to cache block size, which is one stride.

## III. WARP SIZE IMPACT

In this section we report how warp size impacts, the number of idle cycles, memory access coalescing, and performance. We do not report SIMD efficiency our as the activity factor ([8]) shows little variation (less than 1%) under warp sizes studied here. In the interest of space, we focus on a subset of four benchmarks representing different behaviors of the complete set used in this work. See Section V for methodology.

**Memory access coalescing.** Memory accesses made by threads within a warp are coalesced into fewer memory transactions to reduce bandwidth demand. We measure memory access coalescing using the following equation:

$$Coalescing\ rate = \frac{Total\ memory\ insn.}{Total\ off\ chip\ request} \quad (1)$$

Figure 2a compares coalescing rates for different warp sizes. As presented, increasing the warp size improves coalescing rate. An increase in warp size can increase the likeliness of memory accesses made to the same cache block. This increase starts to diminish for warp sizes beyond 32 threads for most benchmarks as coalescing width (16 words of 32-bit) becomes saturated. Accordingly, enlarging the warp beyond a specific size, returns little coalescing gain. Another reason for the little gain is that most workloads implicitly optimize coalescing for conventional 32 threads per warp machines.

**Idle cycles.** Figure 2b reports idle cycle frequency for different warp sizes. Idle cycles are cycles when the scheduler finds no ready warps in the pool. Core idle cycles are partially the result of branch/memory divergences which inactivate otherwise ready threads [11]. Small warps may compensate branch/memory divergence by hiding idle cycles (e.g., MU). On the other hand, for some benchmarks (e.g., BKP), small warps lose many coalescable memory accesses, increasing memory pressure. This pressure increases average core idle durations compared to larger warps (e.g., BKP).

**Performance.** Figure 2c reports performance for different warp sizes. An increase in warp size can have opposite effects on performance. Performance can improve if an increase in memory access coalescing outweighs synchronization overhead. Performance can suffer if the synchronization overhead associated with large warps exceeds coalescing memory access gains. As reported, warp size has significant impact on performance. Performance improves in BKP with warp size. Performance is lost in MU as warp size increases. HSPT performs best under average warp sizes (16 threads). CP is less sensitive to warp size.

We conclude from this section that warp size can impact performance in different ways. We introduce DWR as a solution to achieve high coalescing rate of large warps and low idle cycle of small warps simultaneously.

## IV. DYNAMIC WARP RESIZING

DWR aims at achieving benefits associated with both small and large warps. DWR is a microarchitectural solution that starts with small warps (as wide as SIMD width and hereafter referred to as sub-warps) but adapts to using larger warp sizes upon encountering specific program behaviors. This dynamic increase in warp size increases memory accesses coalescing (often absent from systems using small warps) and relies on using barrier synchronizers to synch and combine multiple sub-warps. DWR schedules sub-warps independently and synchronizes them to execute memory instructions in combined form and in a larger warp. DWR extends the ISA to implement this synchronization and warp scheduler to support warp combining. DWR's architecture is shown in Figure 3.

We present the proposed microarchitecture in subsection IV.A. Deadlock freedom and unnecessary synchronizations are presented in subsections, IV.B and IV.C, respectively. We introduce the operation of the new instruction supporting deadlock freedom and avoiding unnecessary synchronization in subsection IV.D. Finally, we evaluate hardware overhead in subsection IV.E.

### A. Microarchitectue

DWR groups and issues warps with different sizes; large warps are employed for specific instructions leaving sub-warps for other instructions. Partner sub-warps are synchronized to build one large warp to execute the specific instructions. Specific instructions include a group of static low-level PTX instructions [13], referred to as Large-wArp-inTensive instructions or LATs. LAT's main difference from other instructions is that LATs are executed faster under large warps. Non-LATs are always executed using sub-warps. LATs, on the other hand, are executed using large warps built from multiple sub-warps. DWR's warp scheduler combines multiple sub-warps into one large warp upon realizing that all partner sub-warps are ready to execute. A single bit per sub-warp, referred to as the combine-ready status bit, is used to make this decision.

**Synchronization.** Since the scheduler can select sub-warps in any order, some sub-warps may reach specific LATs earlier than other partner sub-warps. To guarantee that all partner sub-warps are ready to execute the associated LAT, we enforce a synchronization barrier just before the LAT. This synchronization can be realized using static or dynamic approaches. *Static synchronization,* which is used in this study, extends the ISA and hardware to support this inter-partner sub-warp synchronization barrier. During compile time, each LAT is replaced by two instructions: 1) a barrier among partner sub-warps and 2) the original LAT. The first instruction (LAT barrier) guarantees that all partner sub-warps have arrived. The second instruction (LAT) is executed using a large warp. Listing 1a shows part of a typical kernel (BFS benchmark) in PTX syntax. 1b shows the transformed code compiled for DWR where the bar.synch_partner is the LAT barrier instruction. Alternatively, *dynamic synchronization* (not used here) can be designed to detect an LAT after decode and synchronizes the partner sub-warps on the instruction in future executions. The dynamic approach keeps DWR binary compatible with the baseline but requires a learning phase before it can identify LAT instructions.

**Selecting LAT.** Using PTX's virtual ISA terminology [13], the candidates for LATs can be load/store from/to global/local/param space or load from const space. These instructions access global memory explicitly. Our baseline architecture is not capable of coalescing memory accesses of const space. Therefore, we consider load/store instructions from/to global/local/param space as LATs.

**Sub-warp Combiner.** Sub-warp Combiner (SCO) is used to construct large warps upon issuing an LAT. The sub-warp synchronizer sends a signal to SCO to identify sub-warps synchronized on an LAT. Sub-warps stay waiting until synchronizer marks them as combine-ready. The combine-ready status shows that all sub-warps have reached the LAT barrier and are ready to be combined and execute the associated LAT. SCO merges active masks of the combine-ready sub-warps, issuing one larger warp. The maximum number of combinable sub-warps (size of the largest warp) is a statically configurable parameter in DWR. A higher maximum provides more opportunities to perform inter-warp memory access coalescing while imposing larger synchronization overhead. In this study we evaluate the following maximum large warp sizes; 2X, 4X and 8X larger than sub-warp size.

### B. Deadlock freedom

The microarchitecture described above may lead to deadlock situation in two cases:

    1) LAT barrier plus another LAT barrier

    2) LAT barrier plus __syncthreads()

Generally, in both deadlock cases, partner sub-warps wait on two or more different barriers preventing uniform barrier freedom. This happens if there is a divergence within a large warp and sub-warps execute different paths and different LATs

```
cvt.u64.s32    %rd1, %r3;           cvt.u64.s32    %rd1, %r3;
ld.param.u64   %rd2, [__parm1];     ld.param.u64   %rd2, [__parm1];
add.u64        %rd3, %rd2, %rd1;    add.u64        %rd3, %rd2, %rd1;
ld.global.s8   %r5, [%rd3+0];       bar.synch_partner    0;
mov.u32        %r6, 0;              ld.global.s8   %r5, [%rd3+0];
setp.eq.s32    %p2, %r5, %r6;       mov.u32        %r6, 0;
@%p2 bra       $Lt_0_5122;          setp.eq.s32    %p2, %r5, %r6;
mov.s16        %rh2, 0;             @%p2 bra       $Lt_0_5122;
st.global.s8   [%rd3+0], %rh2;      mov.s16        %rh2, 0;
(a)                                 bar.synch_partner    0;
                                    st.global.s8   [%rd3+0], %rh2;
                                    (b)
```

**Listing 1. (a) Original PTX instruction sequence of the baseline. (b) DWR-specific generated code supporting inter partner sub-warp synchronization on LATs.**

```
1: if( sub_warp_id == 0){       1: if( sub_warp_id == 0){
2:     regA = gmem[idxA];       2:     regA = gmem[idx];
3: }                            3: }
4: regB = gmem[idxB];           4: __syncthreads();
           (a)                             (b)
```

**Listing 2. Deadlock cases associated with baseline DWR. (a) One of the partner sub-warp waits at LAT barrier #2 and the other sub-warp waits on LAT barrier #4. (b) One of the partner sub-warp waits at LAT barrier #2 and the other sub-warp waits at syncthread #4.**

```
if( warp_id == 0){
    __syncthreads(); // warp-0 is locked here
}else{
    __syncthreads(); // warp-1 is locked here
}
```

**Listing 3. A case with CUDA standard API which is expected to fall in deadlock.**

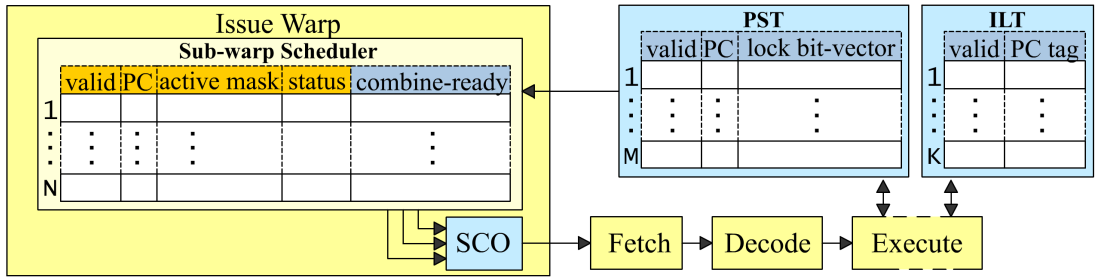

**Figure 3. DWR microarchitecture. Synchronization instruction uses PST and ILT to synchronize sub-warps. SCO issues one large warp when the sub-warps are synchronized. N sub-warps are synchronized in M large warps.**

(or __syncthreads()). Listing 2 presents two high-level CUDA-like examples on how the deadlock can occur under DWR.

These deadlock cases are similar to what could happen under CUDA standard API as shown in Listing 3. However, under Tesla, this does not lead to deadlock as described by Wong et al [17]. Synchronization hardware does not synchronize the threads at specific instructions, it only locks the threads until they reach 1) __syncthreads() or 2) program exit. We solve the baseline's deadlock using the same approach: *LAT insn. barrier does not synchronize threads at specific instructions, it only locks threads until they reach 1) LAT insn. barrier, 2) __syncthreads() or 3) program exit().* Consequently, in both cases presented in Listing 2, deadlock is avoided by releasing both sub-warps. As a result, however, they cannot construct one uniform warp since they have different PCs. Consequently, in this case, partner sub-warps are regrouped in different warps.

### C. Selective synchronization

Synchronizing partner sub-warps in situations like Listing 2 comes with minor coalescing gain and significant synchronization overhead. We refer to this non-performance benefiting synchronization as non-benefiting LAT (or simply NB-LAT) synchronization. NB-LAT synchronization frequently occurs in applications highly prone to branch divergence (like BFS, MU, MP and NQU). Detecting such NB-LAT synchronization instructions statically is not possible since branch divergence occurrence is decided dynamically. We detect NB-LAT synchronization using bar.synch_partner instruction dynamically and as follows. Once the instruction detects that the partner sub-warps are synchronized at different program counters, it stores one of the different PCs in a table (referred to as ignore list table or ILT). ILT stores the PC of NB-LAT synchronizations dynamically and is accessible by only bar.synch_partner. To improve performance, bar.synch_partner does not lock the sub-warp if the bar.synch_partner's PC exists in ILT.

### D. LAT barrier instruction

In this section we discuss the operations of bar.synch_partner. We refer to the group of sub-warps configured statically to be synchronized at LAT as the *partner sub-warp group*. To manage partner sub-warp synchronization, one entry per partner sub-warp group is stored in partner-synch table (PST). Each PST entry consists of the program counter (PC) and a lock bit vector. Bar.synch_partner operates on two inputs: sub-warp identifier and PC. Upon executing bar.synch_partner, if the PC exists in ILT, no further operation is performed. Otherwise, the following operations are performed sequentially in two steps when a sub-warp executes the instruction.

**Step 1. Updating PC, lock bit vector and ILT.** If the group entry is not valid, the entry's PC is updated and the associated bit of the sub-warp in the bit vector is set. If the group entry is valid and its PC is equal to the sub-warp's PC, only the bit vector is updated. If the PC is valid and it is not equal to the barrier instruction PC, the bit vector is updated and the sub-warp's PC is inserted into ILT and will be ignored in future synchronizations on this instruction.

**Table 1. Benchmarks Characteristics. LAT shows the number of LATs and the number of ignored LATs under DWR (with maximum warp size of 64).**

| Name | Abbr. | Grid Size | Block Size | #Insn | LAT |
|---|---|---|---|---|---|
| BFS Graph[2] | BFS | 16x(8) | 16x(512) | 1.4M | 7/15 |
| Back Propagation[2] | BKP | 2x(1,64) | 2x(16,16) | 2.9M | 0/17 |
| Coulumb Poten. [1] | CP | | (16,8) | 113M | 0/5 |
| Dyn_Proc[2] | DYN | 13x(35) | 13x(256) | 64M | 0/9 |
| Gaussian Elimin.[2] | GAS | 48x(3,3) | 48x(16,16) | 9M | 0/11 |
| Hotspot[2] | HSPT | (43,43) | (16,16) | 76M | 0/20 |
| Fast Wal. Trans.[14] | FWAL | 6x(32) 3x(16) (128) | 7x(256) 3x(512) | 11M | 0/7 |
| MUMmer-GPU++[6] | MP | (196) | (256) | 139M | 36/54 |
| Matrix Multiply[14] | MTM | (5,8) | (16,16) | 2.4M | 0/7 |
| MUMmer-GPU[1] | MU | (196) | (256) | 75M | 3/11 |
| Nearest Neighbor[2] | NNC | 4x(938) | 4x(16) | 5.9M | 17/17 |
| N-Queen [1] | NQU | (256) | (96) | 1.2M | 0/10 |
| Scan[14] | SC | (64) | (256) | 3.6M | 0/5 |
| Needleman-Wun. [2] | NW | 2x(1) … 2x(31) (32) | 63x(16) | 12M | 3/26 |

**Step 2. Updating sub-warps status.** If the bit vector is all set, the barrier unlocks all partner sub-warps and marks them as combine-ready in the scheduler. Otherwise, the sub-warp is marked as waiting at synch_partner and stays waiting for other partners.

We assume 24-cycle pipelined latency (equal to the pipeline depth) for performing one bar.synch_partner operation for a sub-warp.

*E. Hardware Overhead*

The baseline warp scheduler updates the status of multiple warps concurrently. SCO combines sub-warps with combine-ready status issuing one large warp. In order to simplify our design, SCO finds combine-ready sub-warps within a limited ID distance. The distance limitation is determined by the pre-decided maximum warp size. For example, if the maximum warp size is four sub-warps, SCO checks sub-warp identifiers between ix4 and (i+1)x4-1. The identified sub-warps are synchronized by LAT barrier.

In order to perform precise operations, warp size should be passed in conjunction with the issued warp (in conjunction with active mask). The warp size (number of sub-warps per warp) can be different multiples of SIMD width (sub-warp size). Knowing the warp size is necessary for the pipeline front-end so it can fetch and decode sub-warps of a large warp. In the pipeline back-end, knowing the sub-warp identifier and the associated active mask is enough to read registers, execute and write-back.

To support ISA extension in the hardware, we assume one entry per large warp in PST. Assuming 8 sub-warps per large warp, each entry has a 1-bit validity, 32-bit PC and 8-bit lock bit vector. For 16 large warps per SM, PST's size is 82 bytes per SM. One comparator is needed to compare the PC entry to the synchronization instruction PC to update ILT. While 11 of the workloads used in this study do not store any PC, ILT size reaches a maximum of 36 entries (in MP). We assume a 32-entry 8-way associative table for ILT, which is indexed by PC's lower two bits. Each entry has a 1-bit validity and 30-bit PC tag. Consequently, ILT size is 124 bytes per SM.

Warp scheduler stores 32-bit PC, 8-bit active mask and 2-bit status per warp [12]. Each entry of warp scheduler is slightly extended to support a 3-bit status instead of 2-bit to support combine-ready. Assuming 64KB register file, 16KB shared memory, and 48 KB D-cache per SM, storage requirement of the PST and ILT impose below 1% overhead per SM.

V. METHODOLOGY

We used GPGPU-sim [1] (version 2.1.1b) cycle-accurate simulator to model a general-purpose GPU-like architecture. We modified GPGPU-sim to model large warps (beyond 32 threads). We modified the tool to model memory coalescing similar to compute compatibility 2.0 devices [15]. Specifically, the modifications are made to carry the warp size in conjunction with warp operands in every pipeline stage. The warp size information is used to coalesce the memory accesses of head sub-warps with the tailing sub-warps in the same warp. With such infrastructure, we have implemented DWR as introduced in section IV. The synchronization instruction is not actually added into the benchmark binary. We model the latency of the synchronization instruction by stalling the sub-warp for 24 cycles (equal to pipeline latency).

For each fixed warp size machine studied in this work we assume a coalescing width as wide as warp size. This is machine as wide as the largest warp size for DWR. We used the configurations described in Section II. Each SM is an 8-wide processor exploiting 48KB L1 data cache (64-way, 12-set) and shares 16KB shared memory among 1024 threads. 16K 32-bit registers per SM are reserved for thread context. 16 SMs provide peak throughput of 332.8 GFLOPS. Six 64-bit wide memory partitions provide memory bandwidth of 76.8 GB/s at dual-data rate.

We used a cache block size of 64 bytes, which is equal to memory transaction chunks. Increasing cache block size (and transaction chunk) to 128 bytes, degrades performance.

We used benchmarks from GPGPU-sim [1], Rodinia [2] and CUDA SDK 2.3 [14]. We also included MUMmerGPU++ [6] third-party sequence alignment program. We use benchmarks exhibiting different behaviors: memory-intensiveness, compute-intensiveness, high and low branch divergence occurrence and with both large and small number of concurrent thread-blocks. Table 1 shows our benchmarks and the summary of their characteristics.

VI. RESULTS

In this section, we evaluate DWR and processors using different fixed warps sizes. DWR has three configurable parameters: ILT size, minimum warp size and maximum warp size. We assume a 32-entry, 8-way cache-like ILT, and a minimum warp size equal to SIMD width. We evaluate 16, 32 and 64 maximum warp sizes notated by DWR-16, DWR-32 and DWR-64, respectively. In Section VI.A we present memory access coalescing. Contribution of idle cycles is reported in VI.B. In Section VI.C we report performance. We present performance sensitivity to L1 D-cache, SIMD width and ILT size in Section VI.D.

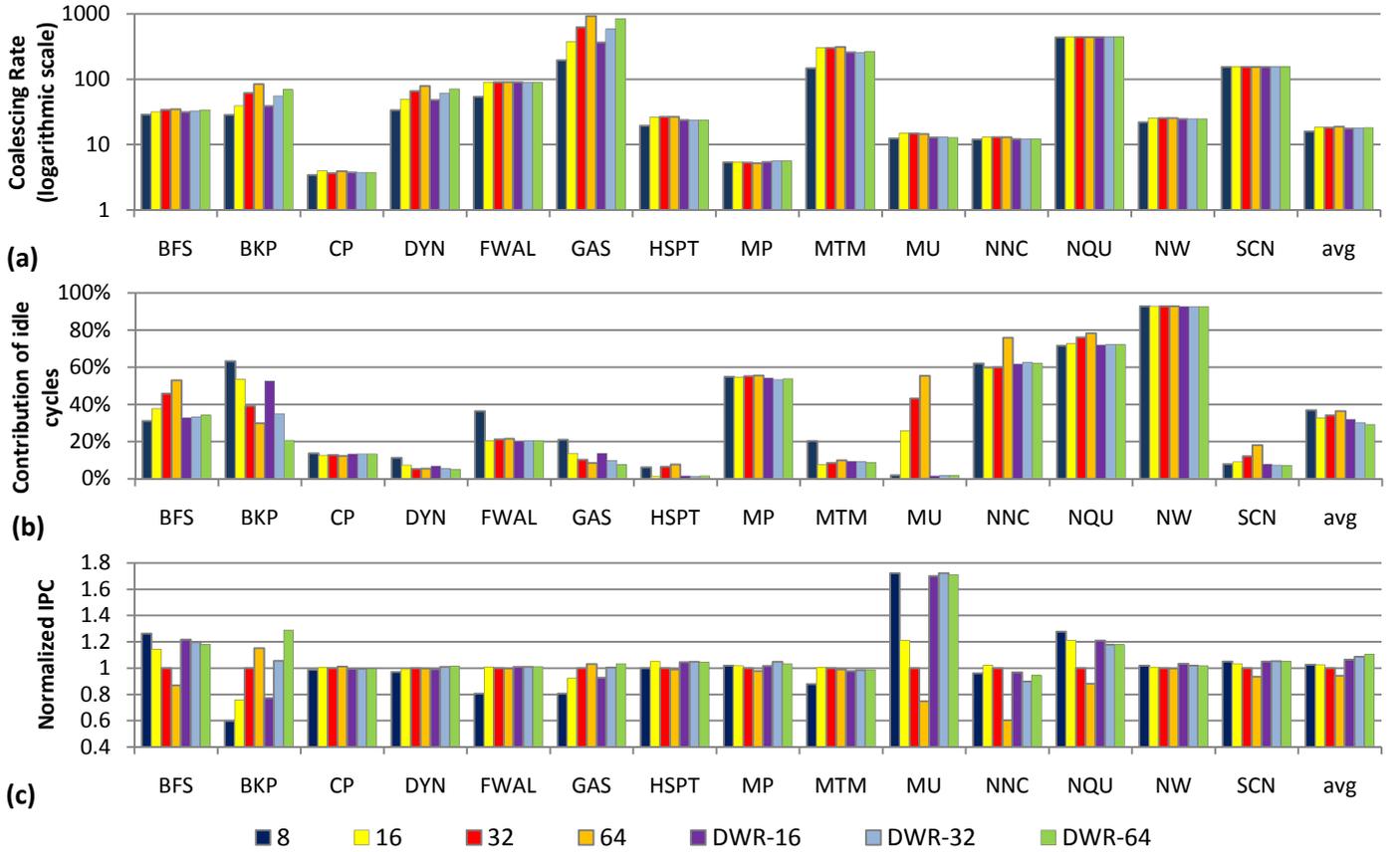

Figure 4. Comparing (a) Coalescing rate, (b) Idle cycle share and (c) Performance for different configurations of DWR and processors using different warp sizes. Each configuration of DWR is notated by DWR-x where x denotes the largest warp size.

### A. Memory access coalescing

Figure 4a reports coalescing rate. As reported, fixed 64-thread per warp provides the highest coalescing rate in most benchmarks. DWR executes most instructions using 8 threads per warps to prevent unnecessary synchronizations. To maintain memory access coalescing of large warps, DWR synchronizes the sub-warps upon memory access. In benchmarks where memory accesses made by neighbor threads is coalescable, DWR provide far higher coalescing rate compared to an 8-thread per warp machine (e.g. BKP, DYN, GAS and MTM). If DWR does not detect any NB-LAT instructions during execution, we expect the coalescing behavior of DWR-X to be similar to using fixed (X) threads per warp machine. However, our estimation of coalescing behavior (i.e., coalescing rate) is affected by cache miss frequency which can depend on sub-warp execution order. Therefore, in benchmarks without NB-LAT (e.g. MTM and FWAL), the coalescing rate of DWR-X and fixed X threads per warp show minor differences due to different warp execution orders under these machines. DWR-64 reaches 97% of the coalescing rate of fixed 64-thread per warp and improves the coalescing rate of fixed 8-thread per warp by 14%.

Under DWR, MU loses coalescing rate considerably compared to fixed large warps. In this benchmark, a considerable part of LATs is placed in the ILT. This coalescing loss, however, does not degrade performance. This is due to the fact that the ILT reduces the synchronization overhead associated with NB-LAT barriers, reducing idle cycles significantly.

### B. Idle cycles

As discussed in Section II, small warps reduce idle cycles by reducing unnecessary waiting due to branch/memory divergence. This idle cycle saving is partially negated since small warps lose memory access coalescing, pressuring the memory subsystem. DWR addresses this drawback by synchronizing sub-warps upon executing memory instructions. DWR reduces unnecessary synchronization of entire warp threads and interleaves sub-warps to hide latency. As reported in 4b, on average, using DWR-64, reduces idle cycles by 26%, 12%, 17% and 25% compared to processors using fixed 8, 16, 32 and 64 threads per warp, respectively. As shown in Figure 4b, DWR-64 shows the lowest average idle cycle share.

Frequent thread synchronization in a block prevents sub-warps from proceeding and hiding each other's latency. For example, MTM unnecessarily synchronizes all threads of a block at every iteration of the main loop. These synchronizations prevent DWR to hide idle cycles effectively across loop iterations using sub-warps.

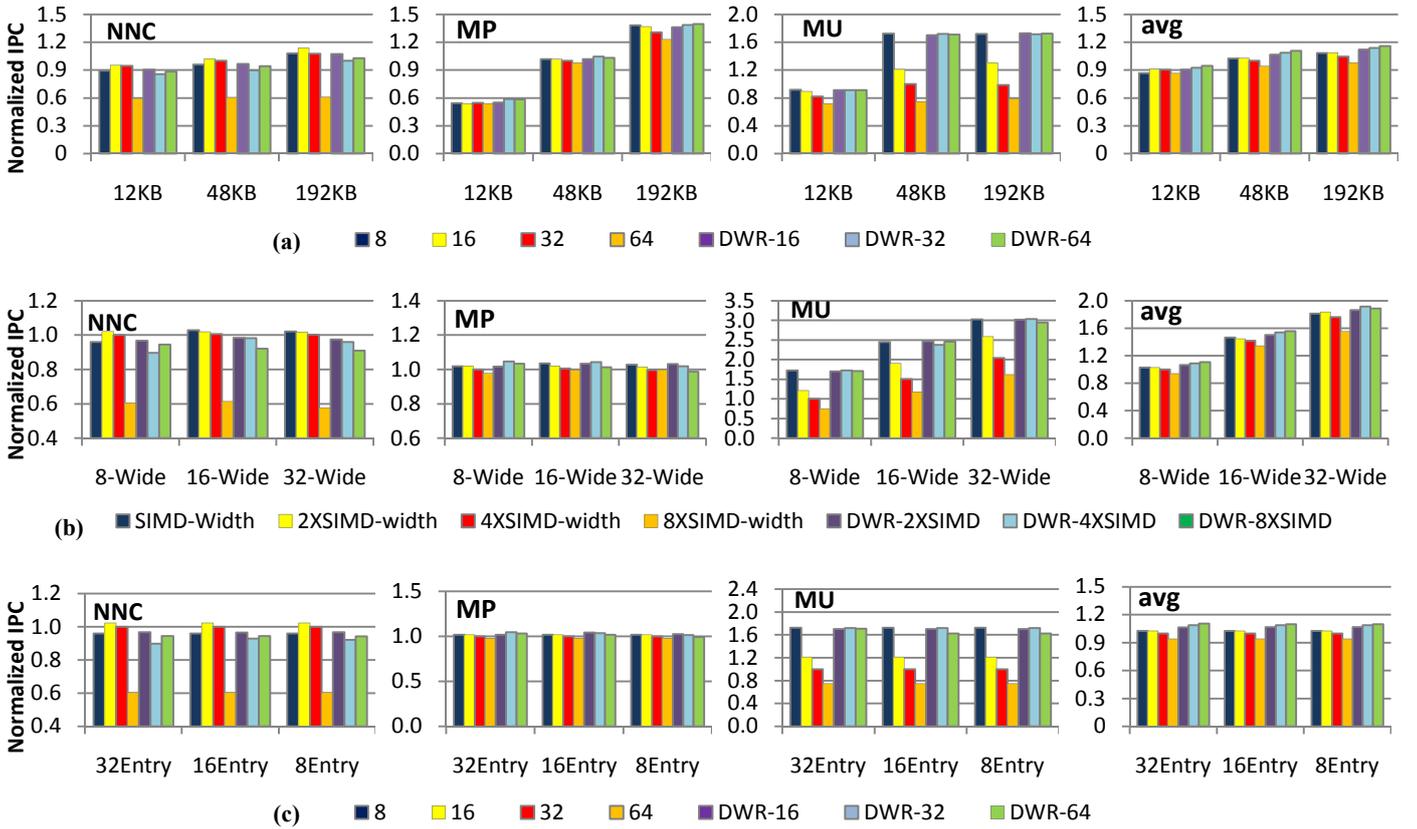

Figure 5. Comparing DWR's performance to GPUs using fixed warp sizes under various configurations. (a) Sensitivity to L1 D-Cache. (b) Sensitivity to SIMD width. For each SIMD width, first four bars from left to right represent machines with fixed warp size. The legend of these bars shows the number of threads per warp (multiples of SIMD width). Last three bars from left represent DWRs with different largest warp sizes. The legend of these bars shows the largest warp size (multiples of SIMD width). (c) Sensitivity to ILT size.

### C. Performance

Figure 4c reports performance for DWR and processors using different fixed warp sizes. In most benchmarks, DWR-64 performs close to the best performing fixed warp size machine. This is due to the fact that DWR combines the benefits of small and large warps. On average, DWR-64 improves performance by 8%, 8%, 11% and 18% compared to fixed 8, 16, 32 and 64 threads per warp machines.

It is important to understand why DWR is outperformed by fixed warp size machines for some applications. NNC, for example, includes 17 LATs in the entire kernel. These instructions are mostly nested at the same nesting level but at different diverging paths. Divergence and sub-warp scheduling order leads to placing the entire 17 LATs into the ILT. Therefore, DWR loses coalescable accesses beyond sub-warp size and performs close to 8-thread per warp machine.

### D. Sensitivity

In this section, we report performance sensitivity to various architectural parameters including L1 D-cache size, SIMD width, and the size of ILT. We limit our report to three representative benchmarks with poor (NNC), average (MP) and good performance (MU) under DWR.

**L1 D-cache.** Baseline architecture uses 48KB (64-set 12-way) L1 cache per SM. Figure 5a reports DWR performance compared to processors using fixed warp sizes and different cache configurations; 4X smaller (12KB, 32-set 6-way) and 4X larger (192KB 128-set 24-way) caches. As reported, employing a smaller cache reduces performance improvements obtainable by DWR. This is due to the following two reasons. First, branch divergence loses its importance as benchmarks become more memory-bound (and less computation-bound) under higher cache miss rate. This reduces branch divergence mitigation benefits of DWR. This explains performance in MU, where even short warps fail to improve performance for small caches. Second, smaller caches reduce memory divergence mitigation benefits of DWR as most cache accesses miss, reducing coalescing opportunities. The gap between best performing fixed warp size and best performing DWR is 8%. Increasing the cache size by 4X affects the gap negligibly and decreasing the cache size by 4X narrows the gab to 4%. Performance improvements achieved for larger caches for DWR can be explained following the same logic.

One important conclusion can be made from the D-cache sensitivity analysis: Large warps are more beneficial when the D-cache is small. This is due to the fact that in systems using small data caches, memory becomes a critical component adding to the importance of memory access coalescing. Notice that under NNC, large warps downgrade performance since NNC's thread-blocks has only 16 threads and large warps underutilize the pipeline.

**SIMD width.** Our baseline architecture exploits 8-wide SMs. Figure 5b, compares DWR and fixed warp size machines under wider SMs; 16-wide and 32-wide. For each SIMD width, the smallest warp size is equal to SIMD width (for DWR). The warp size of each machine is denoted by multiples of SIMD width (warp size in fixed size machine and largest warp size for DWR). Aggressive employment of wide SIMD results in increasing the memory sub-system pressure [12]. Therefore, wider SIMD reduces the impact of branch divergence mitigation benefits of DWR as memory becomes critical. Comparing the best performing DWR to best performing fixed warp size, doubling the SM's SIMD width to 16-lane per SM, reduces the gap to 7%. Further widening the SIMD to 32-lane, reduces this gap to 5%. Notice that NNC and MP show no performance improvements under wider SIMD since NNC uses 16 threads per block and MP is heavily bounded by memory performance.

**ILT size.** In this study, we have assumed 32-entry (4-set 8-wait) ILT. As Figure 5c reports, 2X smaller (16-entry; 4-set 4-way) or 4X smaller (8-entry; 2-set 4-way) table achieve 99% of the performance of the baseline 32-entry table.

## VII. DISCUSSION

In this section we comment on some practical implications and provide more insight.

**Insensitive workloads.** Warp size affects performance in SIMT cores only for workloads suffering from branch/memory divergence or showing potential benefits from memory access coalescing under large warps. Therefore, benchmarks lacking either of these characteristics (e.g. CP and DYN) are insensitive to warp size.

**Enhancing short warps.** DWR can be viewed as a mechanism to enhance performance for GPUs using short warps. Among all configurations, a GPU using 8 threads per warp performs worst for many benchmarks (e.g., BKP) as it suffers from very low memory coalescing. DWR enhances this machine significantly and comes with considerable (up to 116%) returns. However, this machine performs well for computation-bounded benchmarks (e.g. BFS, MU and NQU), which suffer from branch divergence significantly.

**Inter warp memory access coalescing.** DWR can also be used as a mechanism to facilitate inter-warp memory access coalescing. This is achieved by the smallest warp size as the baseline warp size and building larger warps when necessary. DWR combines multiple warps to coalesce memory accesses of the warps.

**Practical issues with small warps.** Pipeline front-end includes the warp scheduler, fetch engine, decode instruction and register read stages. Using fewer threads per warp affects pipeline front-end as it requires a faster clock rate to deliver the needed workload during the same time period. An increase in the clock rate can increase power dissipation in the front-end and impose bandwidth limitation issues on the fetch stage. Moreover, using short warps can impose extra area overhead as the warp scheduler has to select from a larger number of warps. In this study we focus on how warp size impacts performance. The impact of warp size on area and power is part of our ongoing research.

**Register file.** Warp size affects register file design and allocation. GPUs allocate all warp registers in a single row [5]. Such an allocation allows the read stage to read one operand for all threads of a warp by accessing a single register file row. For different warp sizes, the number of registers in a row (row size) varies according to the warp size to preserve accessibility. Row size should be wider for large warps to read the operands of all threads in a single row access and narrower for small warps to prevent unnecessary reading.

**Future generation of GPUs.** The current trend in NVIDIA GPUs indicates a steady growth in the number of threads, warp schedulers, and cores per SM. DWR is designed as a scalable solution and stays effective under an increase in the number of threads/warps per SM. As we presented, wider SIMD limits performance benefits of DWR as it increases the size of the smallest warp size hence imposing higher synchronization overhead. However, the SIMD width of today GPUs (e.g. NVIDIA Kepler [16]) is kept below 16 to prevent design risk [12]. Kepler employs 192 cores per SM and cores are grouped into 12 independent 16-wide SIMD groups. Although we have evaluated DWR under Tesla-like architecture, we believe DWR can improve performance under Fermi and Kepler too.

## VIII. RELATED WORKS

To best of our knowledge, this is the first work investigating warp size issues in GPUs. Kerr et al. [8] introduced several metrics for characterizing GPGPU workloads. Bakhoda et al. [1] evaluated the performance of SIMT accelerators for various configurations including interconnection networks, cache size and DRAM memory controller scheduling. Lashgar and Baniasadi [9] evaluated the performance gap between realistic SIMT cores and semi-ideal GPUs to identify appropriate investment points.

Dasika et al. [4] studied SIMD efficiency according to the SIMD width. Their study shows the frequent occurrence of divergence in the scientific workloads makes wide SIMD organizations inefficient in terms of performance/watt. 32-wide SIMD is found to be the most efficient design for the studied scientific computing workloads.

Jia et al. [7] introduced a regression model relating the GPU performance to microarchitecture parameters such as SIMD width, thread block per core and shared memory size. Their study did not cover warp size but concluded that SIMD width is the most influential parameter among the studied parameter.

## IX. CONCLUSION

In this work we evaluated the performance of Tesla-like GPUs under different warp sizes. We found that small warps serve well for application suffering from branch divergence. On the other hand, large warps are more suitable for memory bounded workloads taking advantage of memory access coalescing.

Based on these findings, we proposed DWR as a dynamic solution aiming at achieving the benefits associated with both large and small warps. Exploring 14 general-purpose benchmarks, DWR outperforms fixed 8, 16, 32 and 64 threads per warp machine up to 2.16X, 1.7X, 1.71X and 2.28X, respectively. Furthermore, our sensitivity analysis shows DWR performs better under narrower SIMD and larger cache.

X. ACKNOWLEDGEMENTS

We should thank Ali Shafiee and anonymous reviewers of ICCD for their valuable comments on this work. This work was partially supported by the School of Computer Science at Institute for Research in Fundamental Sciences (IPM).